\documentclass[letter]{aa} 

\usepackage{graphicx}
\graphicspath{ {./Images/} }
\usepackage[font=small]{caption}
\usepackage{txfonts}
\usepackage{subcaption} 
   
\usepackage{float}
\usepackage{upgreek}
\usepackage{fontawesome5}
\usepackage{ragged2e}
\usepackage{placeins}

\newcommand{\orcid}[1]{\href{https://orcid.org/#1}{\textcolor[HTML]{A6CE39}{\faOrcid}}}

\usepackage[colorlinks=true,linkcolor=blue,citecolor=blue]{hyperref}
\begin{document} 
   \title{Spatially resolved dust properties over 50 kpc in a hyperluminous galaxy merger at $z=4.6$}
   \author{Román Fernández Aranda\inst{1,2}\orcid{0000-0002-7714-688X}
   \and Tanio Díaz Santos\inst{2,3}\orcid{0000-0003-0699-6083}
   \and Evanthia Hatziminaoglou\inst{4,5,6}\orcid{0000-0003-0917-9636}
   \and Manuel Aravena\inst{7}\orcid{0000-0002-6290-3198}
   \and Daniel Stern\inst{8}\orcid{0000-0003-2686-9241} 
   \and Lee Armus\inst{9}\orcid{0000-0003-3498-2973}
   \and Roberto J. Assef\inst{7}\orcid{0000-0002-9508-3667}
    \and Andrew W. Blain\inst{10}\orcid{0000-0001-7489-5167}
    \and Vassilis Charmandaris\inst{1,2,3}\orcid{000-0002-2688-1956}
    \and Roberto Decarli\inst{11}\orcid{0000-0002-2662-8803}
    \and Peter R. M. Eisenhardt\inst{8}
    \and Carl Ferkinhoff\inst{12}
    \and Jorge González-López\inst{13,14}\orcid{0000-0003-3926-1411}
    \and Hyunsung D. Jun\inst{15}\orcid{0000-0003-1470-5901}
    \and Guodong Li\inst{16,17}\orcid{0000-0003-4007-5771}
    \and Mai Liao\inst{7,16,18}\orcid{0000-0002-9137-7019}   
   \and Victoria Shevill\inst{10}
   \and Devika Shobhana\inst{7}\orcid{0000-0002-5033-8056}
   \and Chao-Wei Tsai\inst{16,17,19}\orcid{0000-0002-9390-9672}
    \and Andrey Vayner\inst{9}\orcid{000-0002-0710-3729}
    \and Jingwen Wu\inst{16,17}\orcid{0009-0005-3345-0010}
   \and Dejene Zewdie\inst{7,20,21}\orcid{0000-0003-4293-7507}
    }
    
   \institute{Department of Physics, University of Crete, 70013 Heraklion, Greece 
        \and Institute of Astrophysics, Foundation for Research and Technology - Hellas (FORTH), 70013 Heraklion, Greece 
        \and School of Sciences, European University Cyprus, Diogenes street, Engomi, 1516 Nicosia, Cyprus 
        \and European Southern Observatory, Karl-Schwarzschild-Str. 2, 85748 Garching bei Müchen, Germany 
        \and Instituto de Astrofísica de Canarias, E-38205 La Laguna, Tenerife, Spain 
        \and Universidad de La Laguna, Dpto. Astrofísica, E-38206 La Laguna, Tenerife, Spain 
        \and Instituto de Estudios Astrofísicos, Facultad de Ingeniería y Ciencias, Universidad Diego Portales, Av. Ejército Libertador 441, Santiago, Chile 
        \and Jet Propulsion Laboratory, California Institute of Technology, 4800 Oak Grove Dr., Pasadena, CA 91109, USA
        \and IPAC, California Institute of Technology, 1200 E. California Boulevard, Pasadena, 91125, CA, USA
        \and School of Physics and Astronomy, University of Leicester, Leicester, LE1 7RH, UK
        \and INAF – Osservatorio di Astrofisica e Scienza dello Spazio di Bologna, Via Gobetti 93/3, I-40129 Bologna, Italy
        \and Department of Physics, Winona State University, 175 West Mark Street, Winona, 55987 MN, United States
        \and Instituto de Astrofísica, Facultad de Física, Pontificia Universidad Católica de Chile, Santiago 7820436, Chile
        \and Las Campanas Observatory, Carnegie Institution of Washington,  Raúl Bitrán 1200, La Serena, Chile
        \and Department of Physics, Northwestern College, 101 7th St SW, Orange City, IA 51041, USA
        \and National Astronomical Observatories, Chinese Academy of Sciences, Beijing 100101, China
        \and University of Chinese Academy of Sciences, Beijing 100049, China
        \and Chinese Academy of Sciences South America Center for Astronomy, National Astronomical Observatories, CAS, Beijing, 100101, China
        \and Institute for Frontiers in Astronomy and Astrophysics, Beijing Normal University, Beijing 102206, China
        \and Centre for Space Research, North-West University, Potchefstroom 2520, South Africa
        \and Department of Physics, College of Natural and Computational Science, Debre Berhan University, P.O. Box 445, Debre Berhan, Ethiopia
        }
   \date{Received ...; accepted ...}
 
  \abstract
    {We present spatially resolved dust-continuum ALMA observations from rest-frame $\sim$\,60 to $\sim$\,600~$\upmu$m (bands 3--10) of the hyperluminous hot dust-obscured galaxy (hot DOG) \textit{WISE} J224607.6--052634.9 (W2246--0526), at redshift $z=4.6$. W2246--0526 is interacting with at least three companion galaxies, forming a system connected by tidal streams. We model the multiwavelength ALMA observations of the dust continuum using a modified blackbody, from which we derive the dust properties (mass, emissivity index, area of the emitting region, and temperature) in the hot DOG and resolved structures across a region of nearly $\sim$50~kpc. The peak temperature at the location of the hot DOG, $\sim$\,110~K, is likely the consequence of heating by the central quasar. The dust temperature drops to $\sim$\,40~K at a radius of $\sim$\,8~kpc,
    suggesting that heating by the quasar beyond that distance is nondominant. The dust in the connecting streams between the host and companion galaxies is at temperatures between $30-40$~K, typical of starburst galaxies, suggesting it is most likely heated by recent, in-situ star formation. This is the first time dust properties are spatially resolved over several tens of kpc in a galaxy system beyond Cosmic Noon --this is more than six times the scales previously probed in galaxies at those redshifts.} 

    \titlerunning{Spatially resolved dust properties over 50 kpc in a hyperluminous galaxy merger at $z = 4.6$}
    \authorrunning{R. Fernández Aranda et al.}
   \maketitle
%
\section{Introduction}
\label{sec:intro}

Supermassive black holes (SMBHs) and their host galaxies assembled most of their mass in the early Universe ($z\sim\,1-6$), making it a critical epoch for galaxy evolution. During these periods of intense accretion and growth, SMBHs are expected to be active and obscured by dust and gas \citep[e.g.,][]{hickox2018}, with active galactic nuclei (AGN) feedback likely shaping the properties of their interstellar medium (ISM) \citep[e.g.,][]{fiore2017,lammers2023}. Consequently, the study of obscured quasars at high-redshift is critical for understanding the evolution of massive galaxies, and specifically obtaining spatially resolved observations is key to reveal the effect of AGN on their hosts. 

Spatially resolved studies of dust properties in nearby galaxies are relatively common \citep[e.g.,][]{cooper2012,galametz2012,zhou2016,utomo2019}. At high-redshift, however, the need for high resolution and sensitivity makes them very challenging. The Atacama Large Millimeter/submillimeter Array (ALMA) is one of the few facilities capable of reaching the necessary sensitivity to detect the lowest surface brightness emission of the coldest dust component in the far-infrared (FIR). However, mapping large areas and covering wide dynamic ranges at different frequencies remains very time consuming, even for bright objects. Only a few studies using ALMA have reported spatially resolved dust temperature measurements in high-redshift galaxies. \cite{litke2022} studied a tightly resolved starburst galaxy at $z=5.7$, finding a global temperature of 48.3~K over $\sim$\,4~kpc. \cite{akins2022} presented a $\sim$\,7~kpc resolved temperature map for a star-forming galaxy at $z=7.13$, with a global temperature of $T_d=41^{+17}_{-14}$~K. Among the literature, we identify only three other studies targeting quasar host galaxies with ALMA. \cite{shao2022} observed a FIR luminous quasar at $z=6.0$ and found a clear temperature gradient over $\sim$\,3~kpc radius toward the center, associated with the AGN. The dust temperature peaks at $\sim$\,70~K, and the global temperature is $T_d=53\pm4$~K. \cite{tsukui2023} reported a resolved dust temperature map for a quasar host galaxy at $z=4.4$, with a gradient peaking at $57.1\pm0.3$~K and outer temperatures of $\sim$\,38~K up to $\sim$\,4~kpc from the center. They associate the peak and gradient with warm dust heated by an AGN. Lastly, \cite{walter2022} targeted a quasar at $z=6.9$ and also found a temperature gradient ranging from over 130~K in the central resolution element (200~pc) to 30~K at 500~pc. These studies show how multiband, resolved ALMA observations can probe quasar-heated dust in galaxies at high-redshift.

In particular, hot dust-obscured galaxies \citep[hot DOGs;][]{eisenhardt2012,wu2012}, among all obscured quasar populations, are unique sources to study the interplay between AGN and dust. Given their extreme luminosities \citep[$L_{\mathrm{bol}}>10^{13}\ L_{\odot}$,][]{tsai2015}, strong SMBH accretion \citep{wu2018,li2024}, and high ISM turbulence \citep{diazsantos2021}, the ISM of hot DOGs is likely experiencing strong heating and feedback from the central quasar. In particular, WISE J224607.6--052634.9 (W2246--0526) is the most distant known hot DOG, at $z = 4.6$, only 1.3 Gyr after the Big Bang. Its extreme luminosity \citep[$L_{\mathrm{bol}} = 3.6\times10^{14}\ L_{\odot}$,][]{tsai2018} is mainly driven by its central quasar, which is likely undergoing super-Eddington accretion \citep[$\lambda_{\mathrm{Edd}}=3.6$,][]{tsai2018} and most probably irradiating its surrounding ISM with intense X-ray emission \citep{fernandez2024}. W2246--0526 is also undergoing a multiple merger, with dust tidal streams connecting the hot DOG with at least three minor galaxy companions, as shown by \cite{diaz2018} using ALMA Band 6 dust-continuum observations. The gas kinematics of W2246--0526 strongly suggests the presence of a large-scale turbulent outflow \citep{diaz2016}, as well as several powerful and asymmetric nuclear outflows concentrated within the central $0.22\arcsec$ ($\sim1.5$\,~kpc) (Liao et al. in prep.). 

In this letter, we present spatially resolved dust observations and properties for the entire W2246--0526 merging system over $\sim\,7.7\arcsec$ ($\sim$\,50~kpc). Throughout the paper, we assume a flat $\Lambda$CDM cosmology with $H_0 = 70$ km s$^{-1}$ Mpc$^{-1}$ and $\Omega_{\mathrm{M}}=0.3$. 
\section{Observations and analysis}
\label{sec:obs}

\begin{figure*}[t]
    \centering
        \subfloat{\includegraphics[width=160mm]{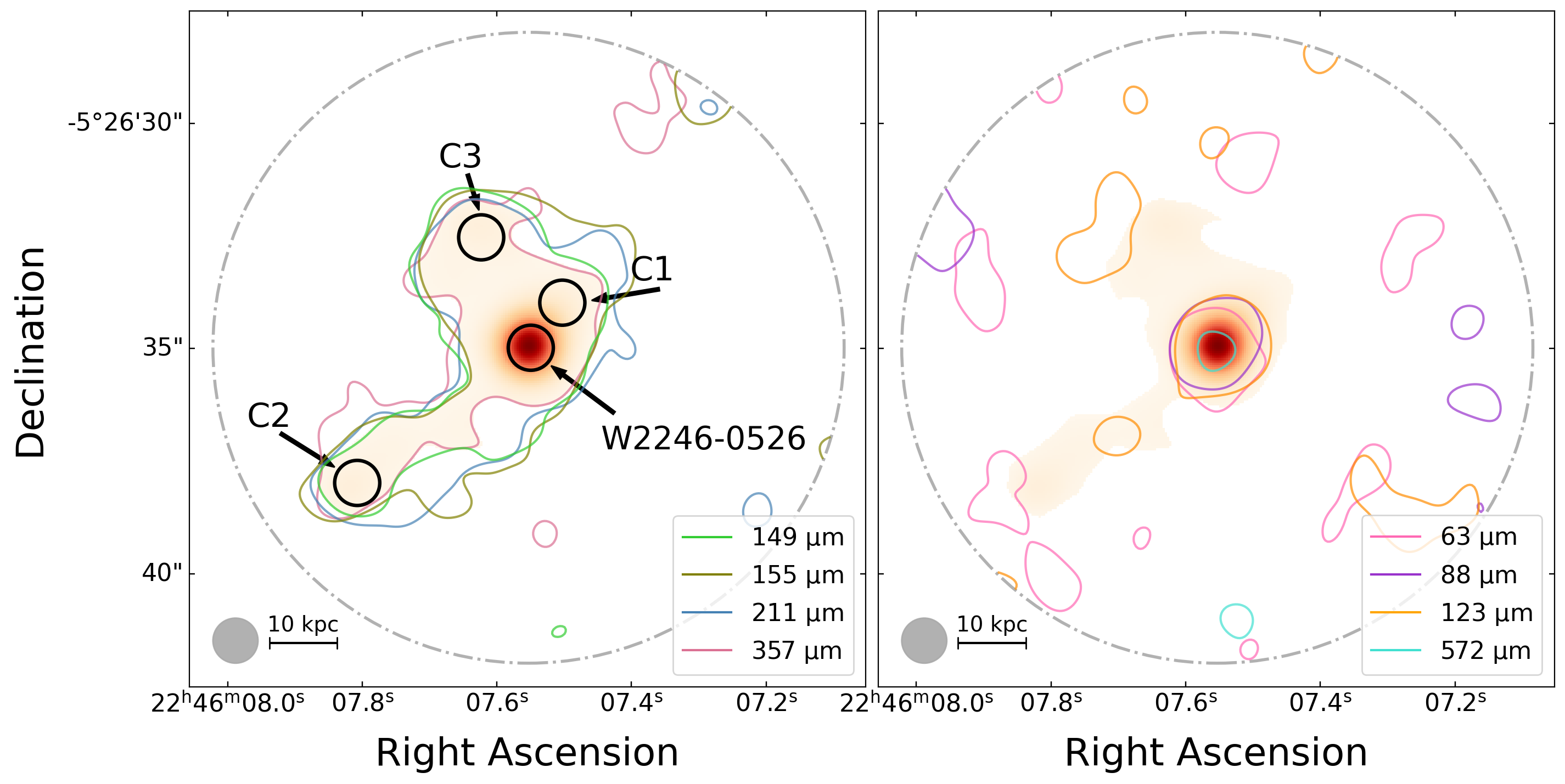}}
        \caption{W2246--0526 continuum emission at various FIR wavelengths used in this study. Shown are the 2.5$\upsigma$ contours for the four spatially resolved datasets (left), along with the three companions detected and labeled in \cite{diaz2018}, and for the four datasets that do not show the entire resolved dust structures (right). The legends indicate their central rest-frame wavelength. The orange background colormap corresponds to the overlapping of the four resolved datasets. The dot-dashed gray circle shows the 7$\arcsec$ radius mask used to clean the datasets. The beam and physical scale are shown in the lower left corner. There is emission from all eight datasets only at the hot DOG position, in the central beam. Note the $\sim35$~kpc dusty stream between W2246--0526 and the companion C2.}
        \label{fig:regions}
\end{figure*}

The W2246--0526 merger system was observed with the ALMA 12-m array during multiple cycles. The observations we compile in this letter are detailed in Table~\ref{tab:Obs}. We reduced and calibrated the data using the Common Astronomy Software Applications \citep[CASA;][]{mcmullin2007} with the standard data reduction procedures. To image the continuum using the \texttt{tclean} task of CASA v6.5, we ran the Hogbom cleaning algorithm \citep{1974A&AS...15..417H} to a flux density threshold of two times the root mean square (rms) of each cube, with the Briggs weighting mode set to robustness = 2.0 (close to natural weighting), and with a uv-taper imaging weight equal to 0.8$\arcsec$ to enhance any extended emission. We further chose to homogenize all the datasets to the same resolution, matching the largest beam after the uv-tapering, by setting the restoring beam in \texttt{tclean} to a common circular beam of FWHM=1.0$\arcsec$. The cleaning was done using a circular mask with a radius of 7$\arcsec$, enough to cover the entire merging system. In each dataset, all emission line-free channels were combined to create the continuum intensity maps, which are shown in the Appendix~\ref{appendix} (Fig.~\ref{fig:A_mom0}). The entire resolved dust structures of the merger presented by \cite{diaz2018} are detected in four of the eight datasets used in this work, and are shown with the overlapped 2.5$\upsigma$ contours in Fig.~\ref{fig:regions}. The non-detection of the extended structures in the other four datasets may be due to smaller maximum recoverable scales in those observations, as shown in Table~\ref{tab:Obs}.

\begin{table*}
    \centering
    \caption{Details of the continuum ALMA observations.}
    \begin{tabular}{*{7}{c}}
    \hline %
    \hline \noalign {\smallskip}
    Central wavelength & Time on-source & Cont. depth & Native beam & MRS & ALMA Band & Project code \\
    $[\upmu\mathrm{m}]$ & [min] & [mJy beam$^{-1}$] & $[\arcsec]$ & $[\arcsec]$ & \\
    (1) & (2) & (3) & (4) & (5) & (6) & (7) \\
    \hline \noalign {\smallskip}
    63 & 30 & 1.15 & $0.39\times0.32$ & 2.9 & 10 & 2017.1.00899.S \\
    88 & 172 & 1.36 & $0.36\times0.31$ & 3.4 & 9 & 2021.1.00726.S \\
    123 & 30 & 0.30 & $0.53\times0.38$ & 3.5 & 8 & 2016.1.00668.S \\
    149 & 124 & 0.040 & $0.44\times0.35$ & 4.5 & 7 & 2017.1.00899.S \\
    155 & 380 & 0.044 & $0.36\times0.30$ & 4.4 & 7 & 2018.1.00333.S \\
    211 & 148 & 0.017 & $0.54\times0.46$ & 5.1 & 6 & 2015.1.00883.S \\
    357 & 293 & 0.007 & $0.50\times0.39$ & 6.1 & 4 & 2019.1.00219.S \\
    572 & 31 & 0.019 & $0.35\times0.30$ & 4.0 & 3 & 2018.1.00119.S \\
     \hline
    \end{tabular} \\
    \vspace{1mm}
    \small
    {\RaggedRight Notes: (1)~Central rest-frame wavelength; (2)~Integration time on-source; (3)~Root mean square of the collapsed dataset; (4)~Average size (FWHM) of the native beam, without uv-tapering; (5)~Maximum recoverable scale; (6)~ALMA receiver band; (7)~ALMA project code\par}

    \label{tab:Obs}
\end{table*}

To characterize the FIR continuum emission in the system, we use a modified blackbody. Starting with the full radiative transfer equation, the flux density observed at frequency $\nu_{\mathrm{obs}}$ is \begin{equation}
    S_{\nu_{\mathrm{obs}}} = \frac{A}{D_L^2}[B_{\nu_{\mathrm{rest}}}(T_d)-B_{\nu_{\mathrm{rest}}}(T_{\mathrm{CMB}})][1-exp(-\tau_{\nu_{\mathrm{rest}}})](1+z)\ ,
    \label{eq:1}
\end{equation} with $A$ being the physical area of emission, $D_L$ the luminosity distance, $T_d$ the temperature of the dust, $T_{\mathrm{CMB}}$ the cosmic microwave background (CMB) temperature at the redshift of the source, and with the blackbody emission being \begin{equation}
    B_{\nu}(T) = \frac{2h\nu^3}{c^2}\frac{1}{exp(\frac{h\nu}{k_B T})-1}\ ,
    \label{eq:2}
\end{equation} and the optical depth of the dust being
\begin{equation}
    \tau_{\nu} = \kappa_0 \left(\frac{\nu}{\nu_0}\right)^{\beta} \frac{M_d}{A}\ ,
    \label{eq:3}
\end{equation} where $\beta$ is the dust emissivity index, $M_d$ is the dust mass, and $\kappa_0$ is the dust opacity at the reference frequency $\nu_0$. Following \cite{scoville2014} and assuming a gas-to-dust ratio of $\delta_{GDR}=200$, given that the metallicity in the hot DOG is $\sim0.5$~Z$_{\odot}$ \citep{fernandez2024}, we adopt $\kappa_0=\kappa_{850\upmu\mathrm{m}}=0.0968\ \mathrm{kg}^{-1} \mathrm{m}^2$ and $\nu_0 = 352.7$ GHz (850$\upmu\mathrm{m}$). We note that $T_d$ is the dust temperature including the CMB heating, which at our redshift and for temperatures above 30~K (see Sect.~\ref{sec:results}) is negligible \citep[less than 1\%;][]{dacunha2013}. 

We fit the modified blackbody of Eq.~\ref{eq:1} to every pixel with emission from the four spatially resolved datasets to derive the properties of the dust in the resolved structures. We use a Markov chain Monte Carlo approach using the \texttt{emcee} \citep{foreman2013} Python package, allowing $A$, $T_d$, $M_d$, and $\beta$ to vary freely using flat priors of $10^{-2}<A<33$~kpc$^2$, $12.6<T_d<300$~K, $10^6<M_d<10^{10}$~$M_{\odot}$, and $1<\beta<3$. The beam area limits the physical area of emission $A$, and the lower temperature limit corresponds to $T_{\mathrm{CMB}}$ at $z=4.601$. We also fit Eq.~\ref{eq:1} for the integrated central beam, corresponding to the hot DOG, for which we have emission from all eight datasets.
\section{Results}
\label{sec:results}
\begin{figure*}[h!]
    \centering
        \subfloat{\includegraphics[width=85mm]{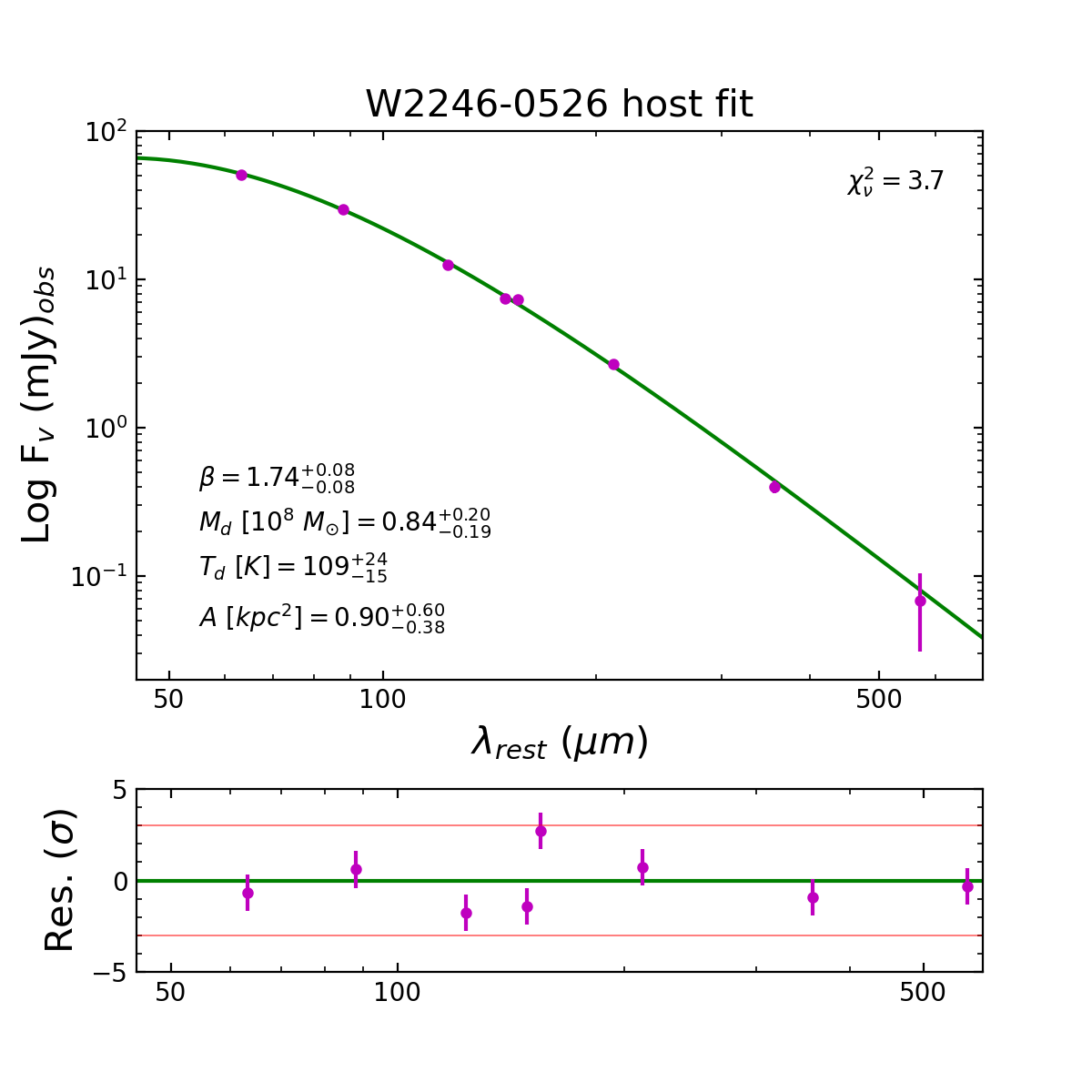}}
        \subfloat{\includegraphics[width=105mm]{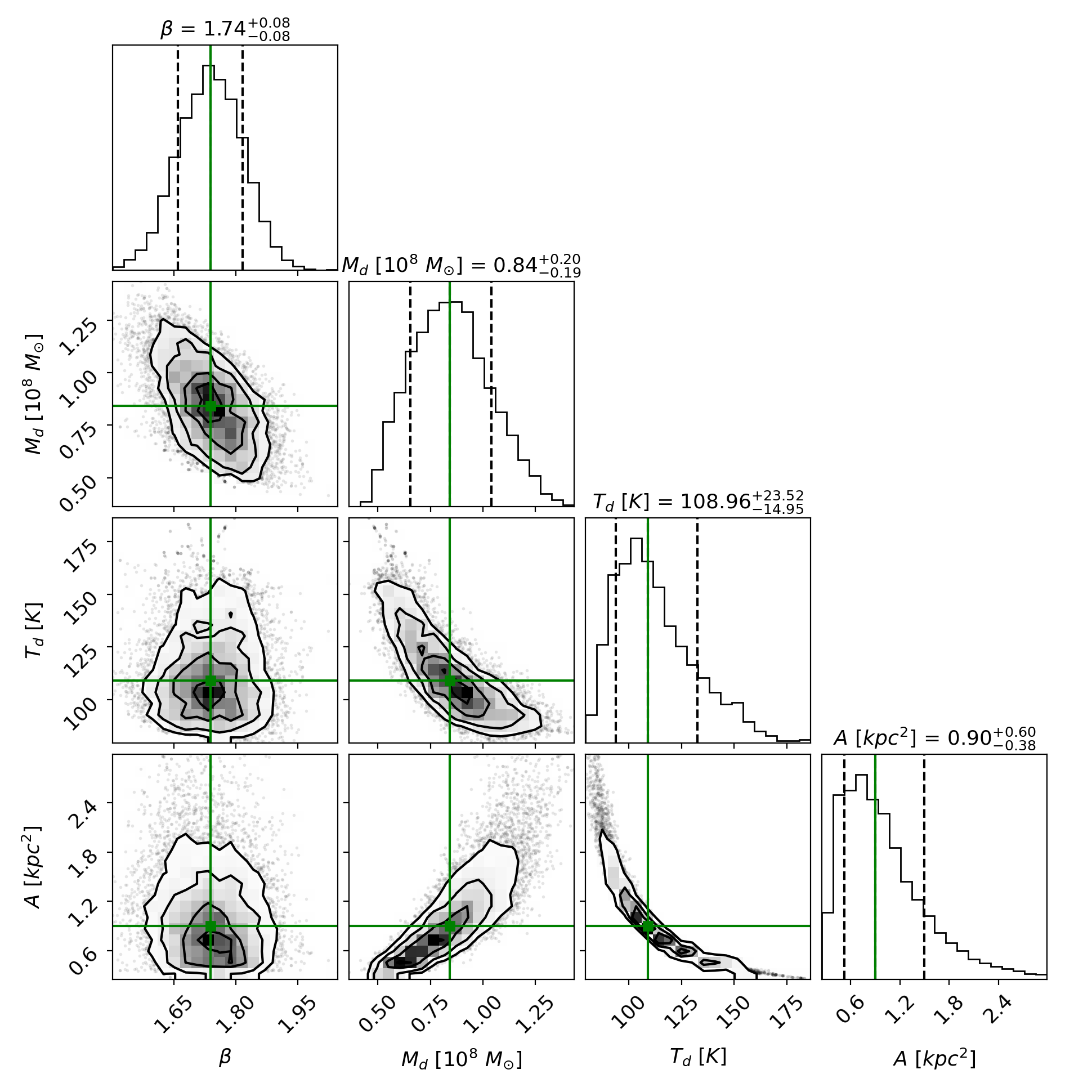}}
        \caption{Dust SED modeling for the central beam (the hot DOG). Left: observed continua flux (purple points) as a function of rest-frame wavelength, and fitted modified blackbody (green line). Best fit parameters and reduced $\chi^2$ of the fit are annotated in the plot, and the residuals of the fit are shown in the bottom panel. Right: corner plot showing the marginalized 1D and 2D posterior probability distributions for the parameters used to fit the modified blackbody. The green solid lines show the position of the best-fit values, and the dashed lines indicate the 68$\%$ confidence interval. The high dust temperature points toward the quasar likely being the primary source of heating.}
        \label{fig:SED}
\end{figure*}

In Fig.~\ref{fig:SED}, we show the spectral energy distribution (SED) fit and the marginalized posterior distributions (PDFs) of the parameters for the integrated central beam, which contains the hot DOG, using all eight datasets. The corresponding fluxes are presented in Table~\ref{tab:Fluxes} and the best fit dust parameters in Table~\ref{tab:Params}. The model parameters for the central beam are well constrained for a single modified blackbody, with $\beta=1.74\pm0.08$, $M_d=0.84_{-0.19}^{+0.20}\times10^8 M_{\odot}$, $T_d=109_{-15}^{+24}$~K, and $A=0.90_{-0.38}^{+0.60}$~kpc$^2$. The best-fit physical area of emission, $A$, is equivalent to that of a circle of diameter $\sim$\,1~kpc, or 0.15$\arcsec$, much smaller than the resolution of our observations (typically 0.4$\arcsec$, see Table~\ref{tab:Obs}) and consistent with a compact heating source. The area obtained from the fit is also consistent with the size of the region where the brightest fine-structure FIR emission lines are produced, with an upper limit on the radius of 800~pc from the central quasar \citep{fernandez2024}. Also, our result for the temperature of the hot DOG is close to that obtained from the W2246--0526 SED fit using UV to submillimeter maps performed by \cite{sun2024}, which found $T_d\sim$\,92~K, and higher than the estimates for the coldest dust component from \cite{diaz2018} and Tsai et al. (in prep.) of $\sim$\,70~K, which assume optically thin dust. 

\begin{table}
    \centering
    \caption{Central beam fluxes.}
    \begin{tabular}{*{3}{c}}
    \hline
    \hline \noalign {\smallskip}
    Central wavelength & ALMA Band & Flux density \\
    $[\upmu\mathrm{m}]$ &  & [mJy] \\
    \hline \noalign {\smallskip}
    63  & 10  & $51.0\pm0.8$ \\
    88  & 9 & $29.6\pm0.6$ \\
    123 & 8  & $12.5\pm0.3$ \\
    149 & 7 & $7.4\pm0.2$ \\
    155 & 7 & $7.3\pm0.2$ \\
    211 & 6 & $2.7\pm0.1$ \\
    357 & 4 & $0.40\pm0.04$ \\
    572 & 3  & $0.068\pm0.037$ \\
    \hline
    \end{tabular} \\
    \vspace{1mm}
    \small
    \label{tab:Fluxes}
\end{table}

\begin{table}
    \centering
    \caption{Dust parameters for the hot DOG.}
    \begin{tabular}{*{3}{c}}
    \hline
    \hline \smallskip
    Parameter & Value & Unit \\
    \hline \noalign{\smallskip}
    $\beta$ & $1.74 \pm 0.08$ & ... \smallskip\\
    $M_d$ & $0.84_{-0.19}^{+0.20}\times10^8$ & $M_{\odot}$ \smallskip\\
    $T_d$ & $109_{-15}^{+24}$ & K \smallskip\\
    $A$ & $0.90_{-0.38}^{+0.60}$ & kpc$^2$ \smallskip\\
    \hline \noalign{\smallskip}
    $\tau_{158\mathrm{\upmu m}}$ & $0.39_{-0.18}^{+0.26}$ & ... \smallskip\\
    $L_{3-1100\mathrm{\upmu m}}$ & $6.4\times10^{13}$ & $L_\odot$ \\
    \hline
    \end{tabular} \\
    \vspace{1mm}
    \small
    {\RaggedRight Notes: $\beta$, $M_d$, $T_d$, and $A$ are the best fit parameters for the central beam. $\tau_{158\mathrm{\upmu m}}$ is estimated using Eq.~\ref{eq:3}, and $L_{3-1100\mathrm{\upmu m}}$ is the integral of the best fit modified blackbody. ~\par}
    \label{tab:Params}
\end{table}

\begin{figure*}[h!]
    \centering
        \hspace{-4mm}
        \subfloat{\includegraphics[width=64.5mm]{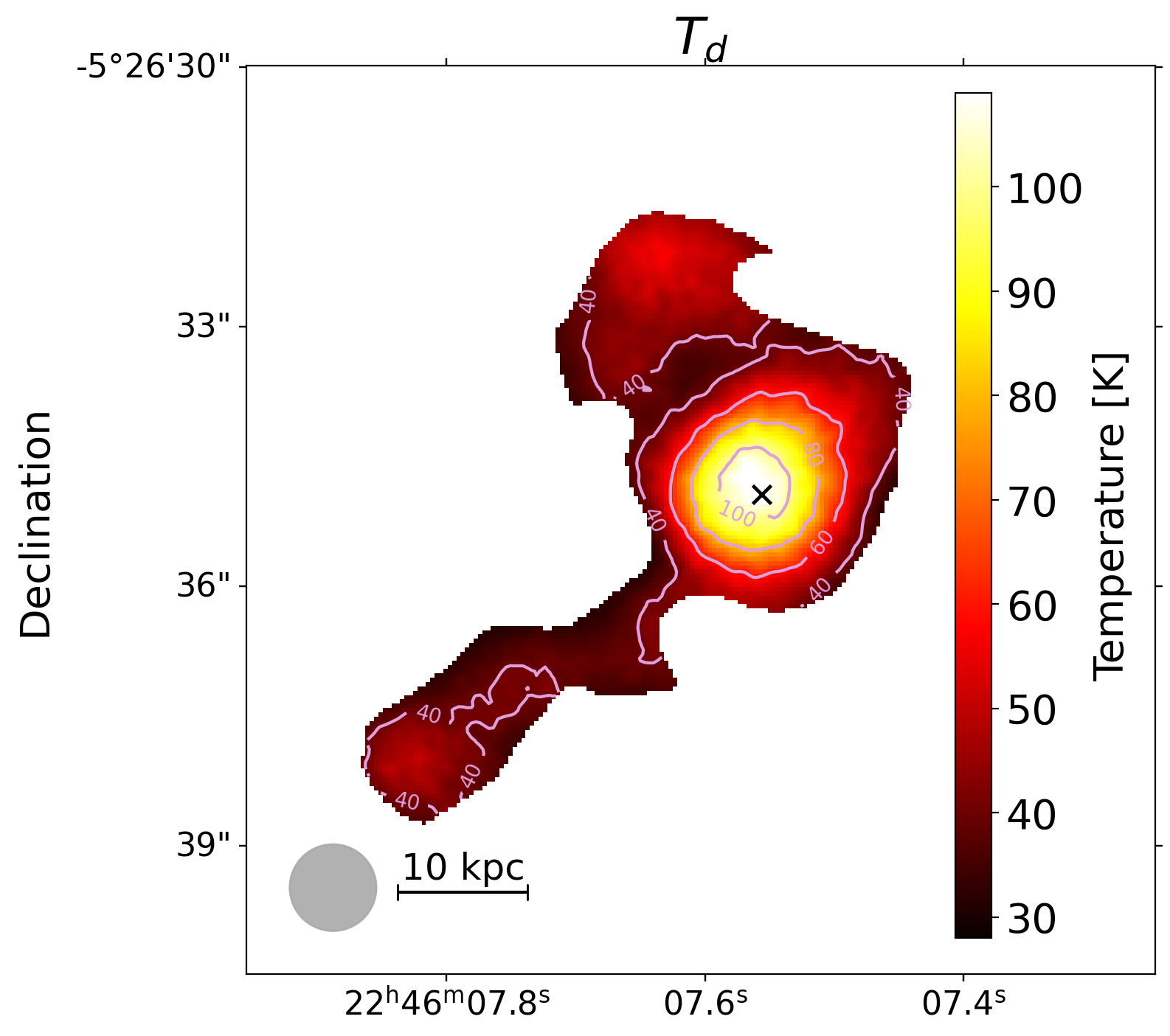}}
        \subfloat{\includegraphics[width=61mm]{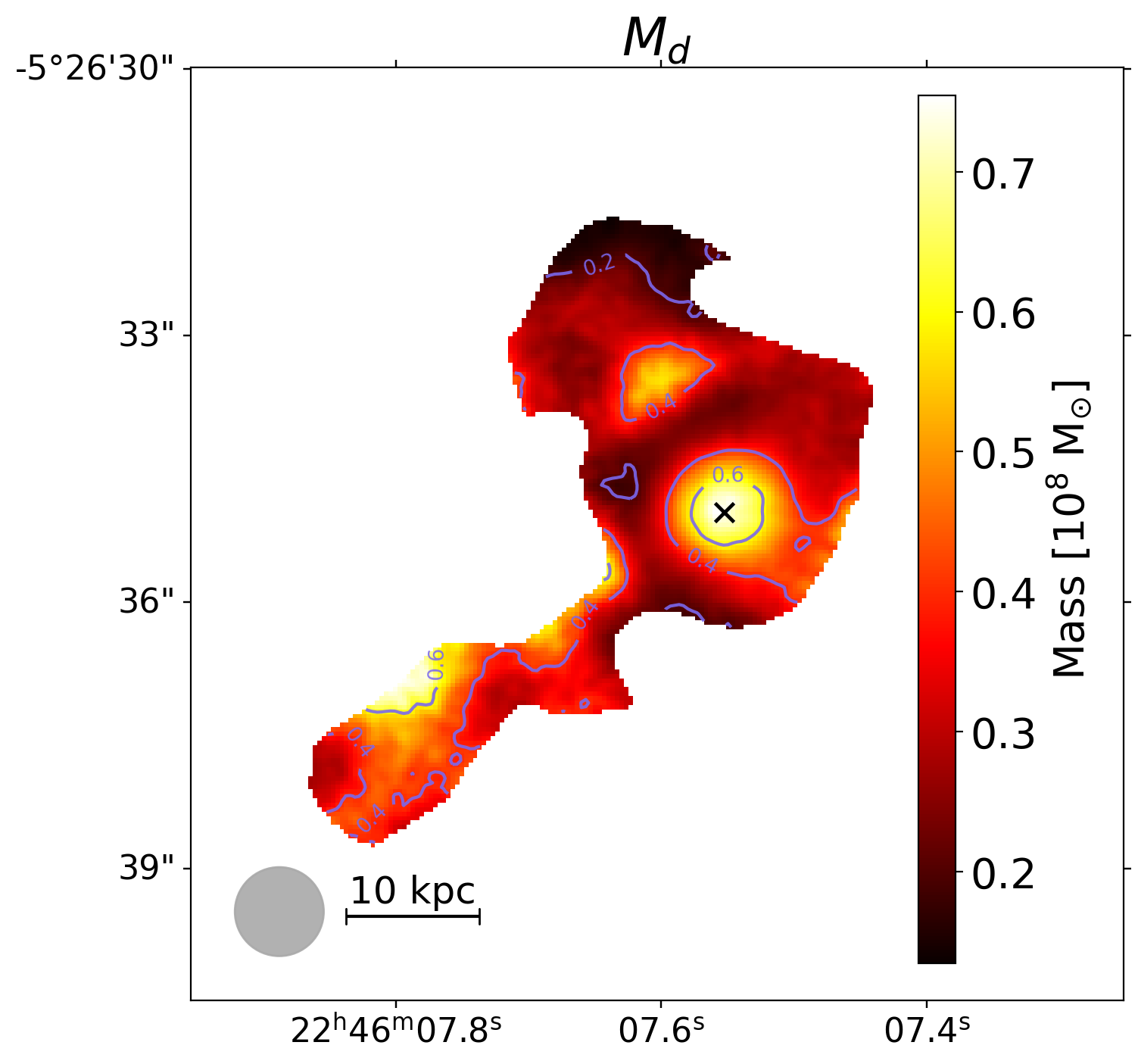}}
        \vspace{0.1cm}
        
        \hspace{-4mm}
        \subfloat{\includegraphics[width=64.5mm]{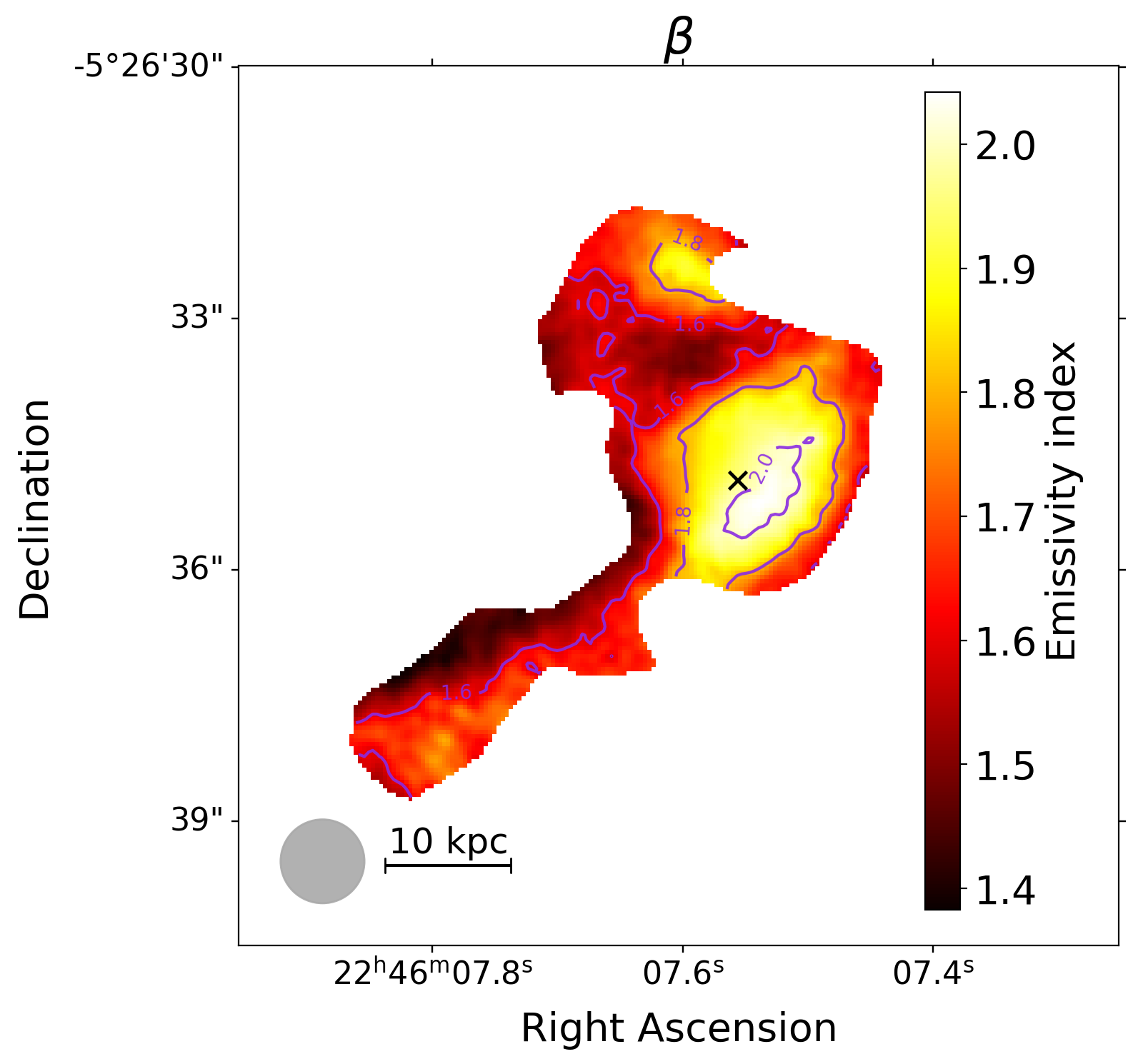}}        \subfloat{\includegraphics[width=61mm]{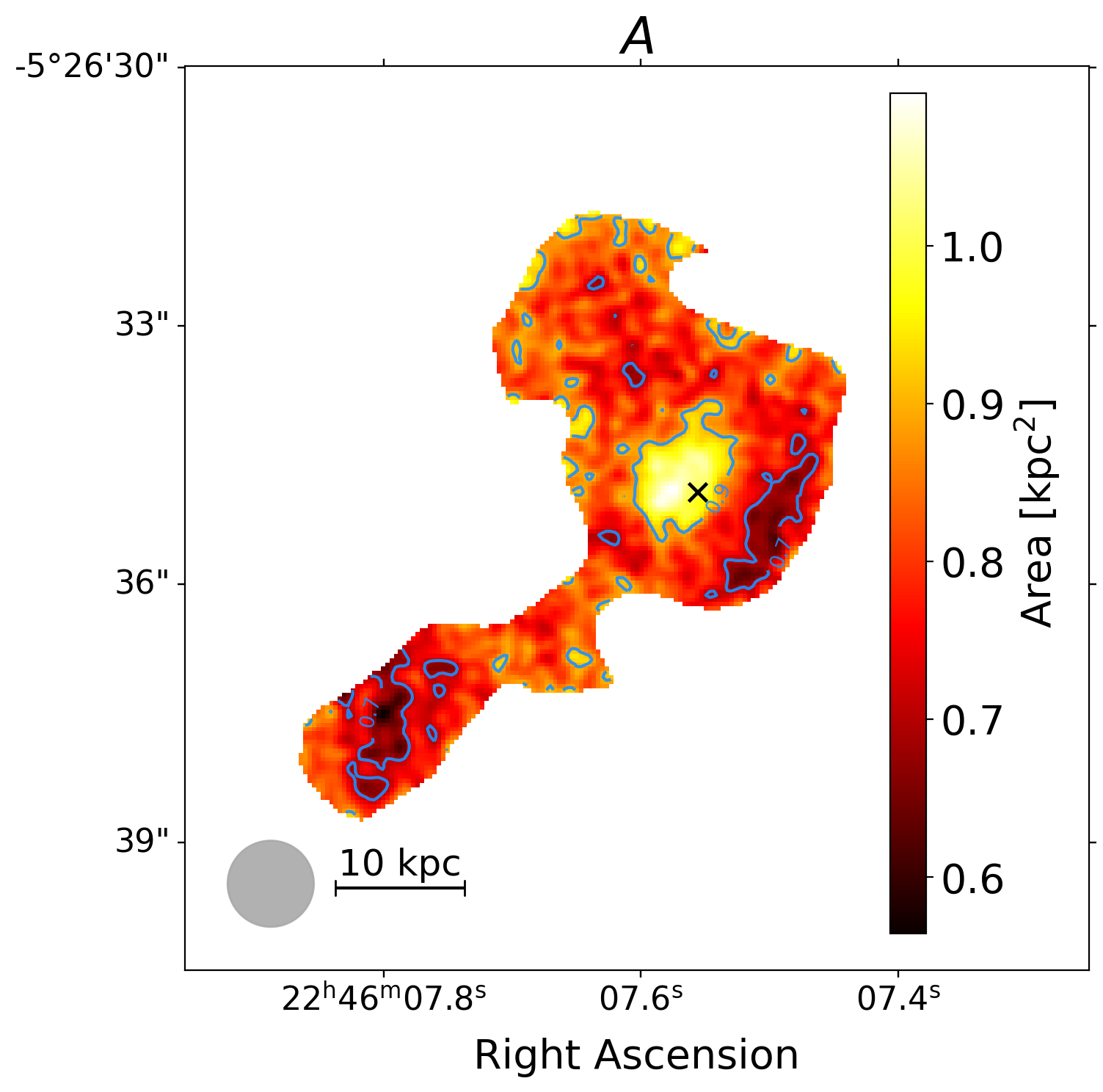}}

        \caption{Resolved maps of the four fitted dust parameters (temperature, mass, emissivity index, and area of the emitting region). All units are per beam. A black cross marks the peak of continuum emission, i.e., the position of the quasar. The beam and physical scale are shown in the bottom-left of each map. The dust properties are derived in the entire merger system, over $\sim$\,50~kpc.}
        \label{fig:dustmaps}
\end{figure*}

In Fig.~\ref{fig:dustmaps} we present the spatially resolved maps of the best-fit parameters using the four datasets where extended emission is detected, from rest-frame 149 to 357~$\upmu$m. The temperature peaks at more than 100~K at the location of the quasar, and gradually descends to $\sim$\,40~K over a scale of $\sim$\,8~kpc, suggesting the quasar is heating the surrounding dust out to this distance. The dust temperature measured in the extended structures surrounding the hot DOG is relatively low, $\sim$\,30--40~K, and typical of star-forming galaxies at similar redshifts \citep[e.g.,][]{bethermin2020,cortzen2020,faisst2020}. The dust mass distribution also peaks in the central region, and the combined dust mass of the whole system is $5.1\times10^8~M_{\odot}$. The dust emissivity index $\beta$ takes values between $\sim$\,1.4--2.0. The values for $M_d$ and $\beta$ are consistent with those from recent studies of various galaxy populations such as quasars, star-forming and submillimeter galaxies at $1<z<8$ \citep[e.g.,][]{dacunha2021,kaasinen2024}. The physical areas of emission vary between $\sim$\,500~pc$^2$ and $\sim$\,1~kpc$^2$, much smaller than the $\sim$\,30~kpc$^2$ beam, implying areal filling factors of dust of 3$\%$ or smaller.  

\begin{figure*}[h!]
    \centering        
        \hspace{-4mm}
        \subfloat{\includegraphics[width=64.5mm]{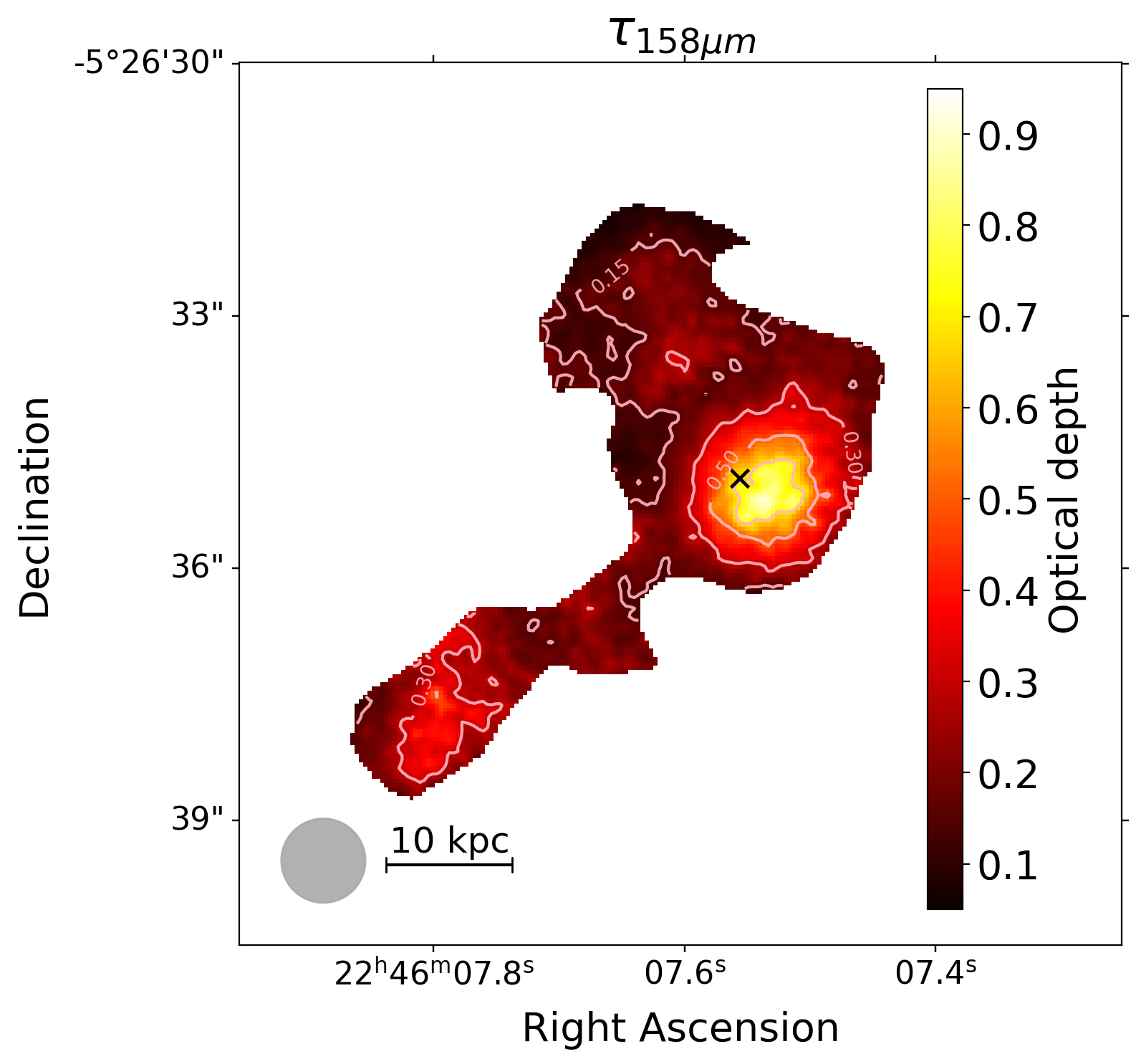}}
        \subfloat{\includegraphics[width=61mm]{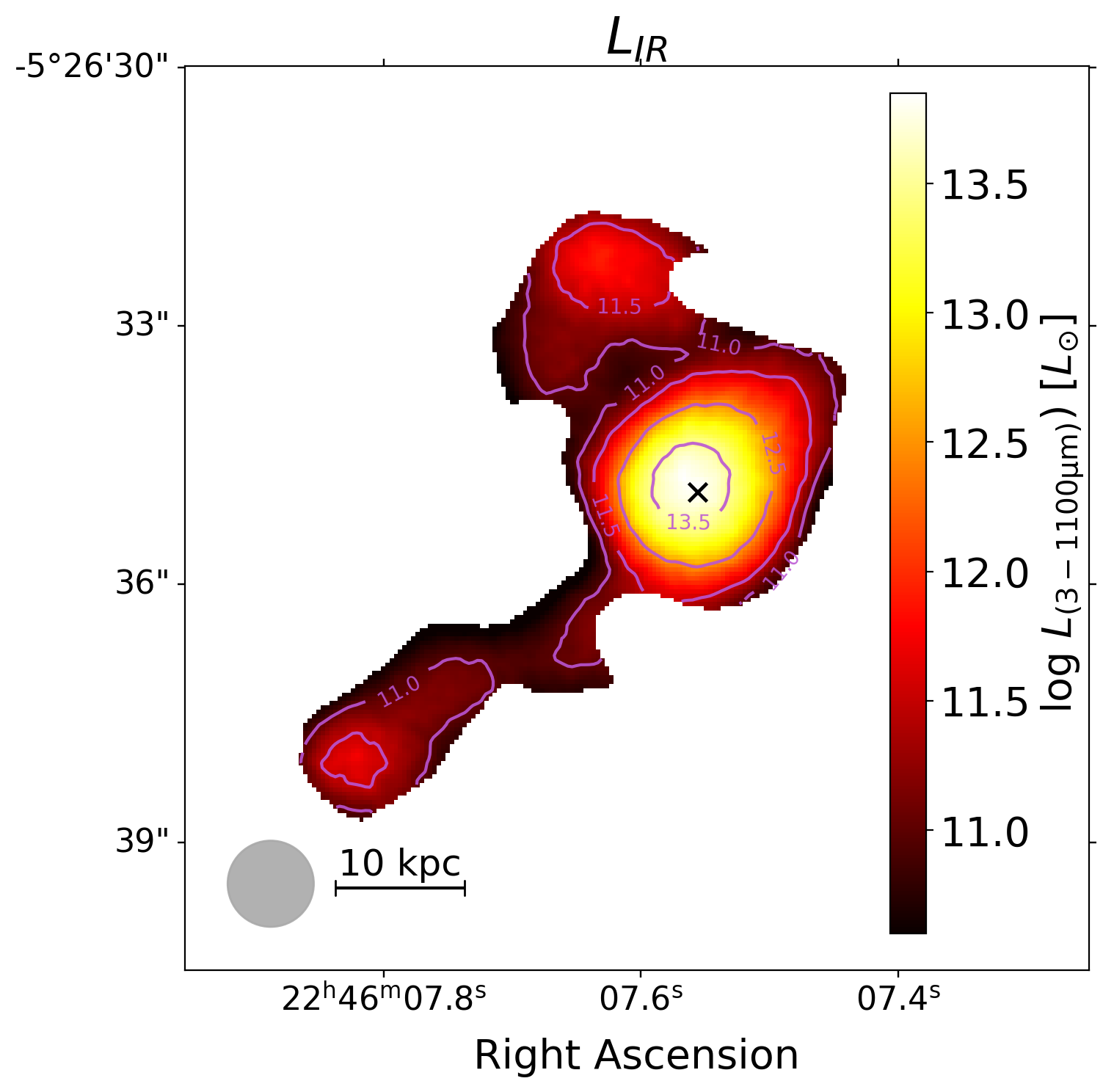}}
        \vspace{0.1cm}     
        \hspace{-4mm}   
        \subfloat{\includegraphics[width=61mm]{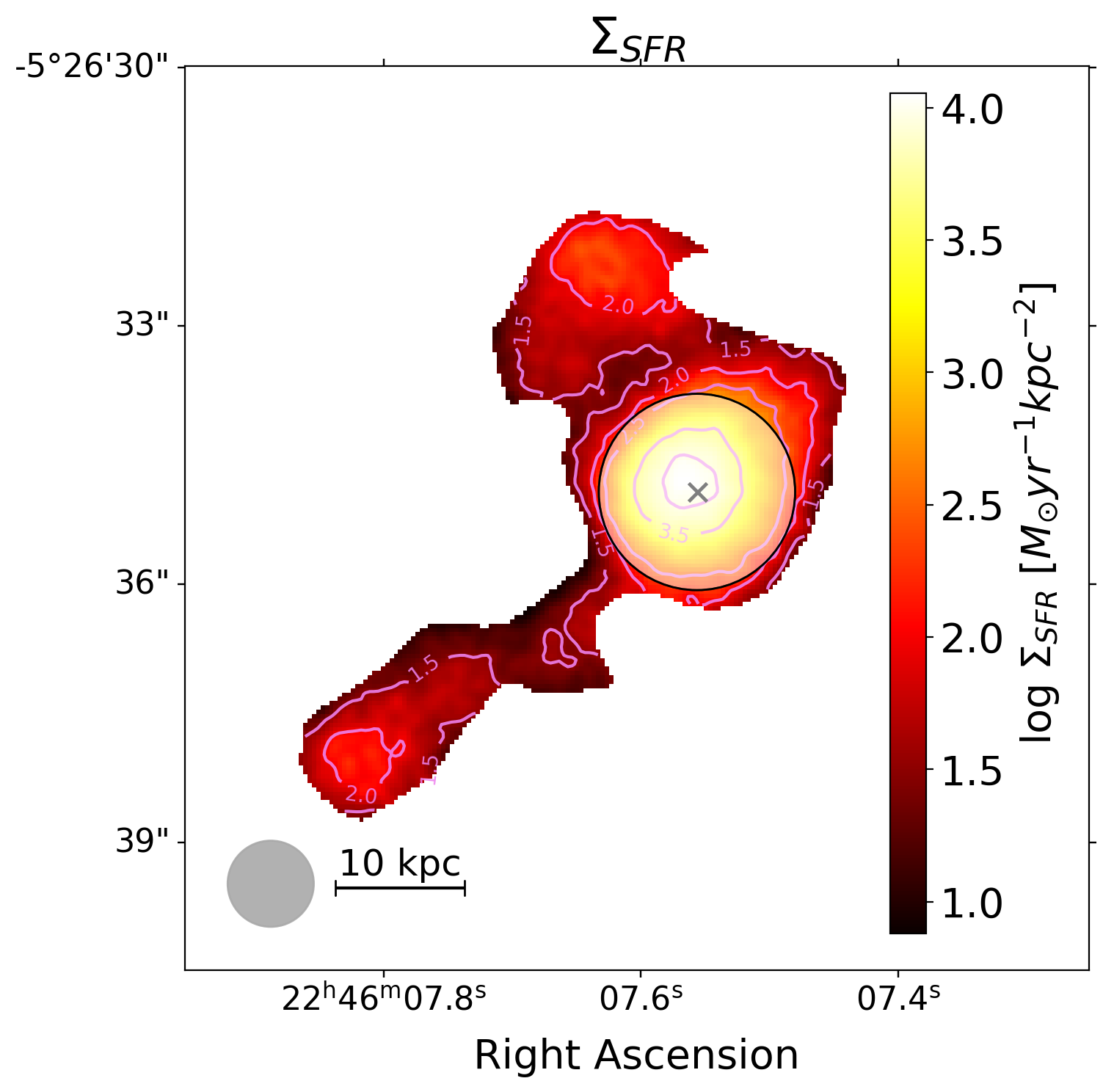}}
        \subfloat{\includegraphics[width=61mm]{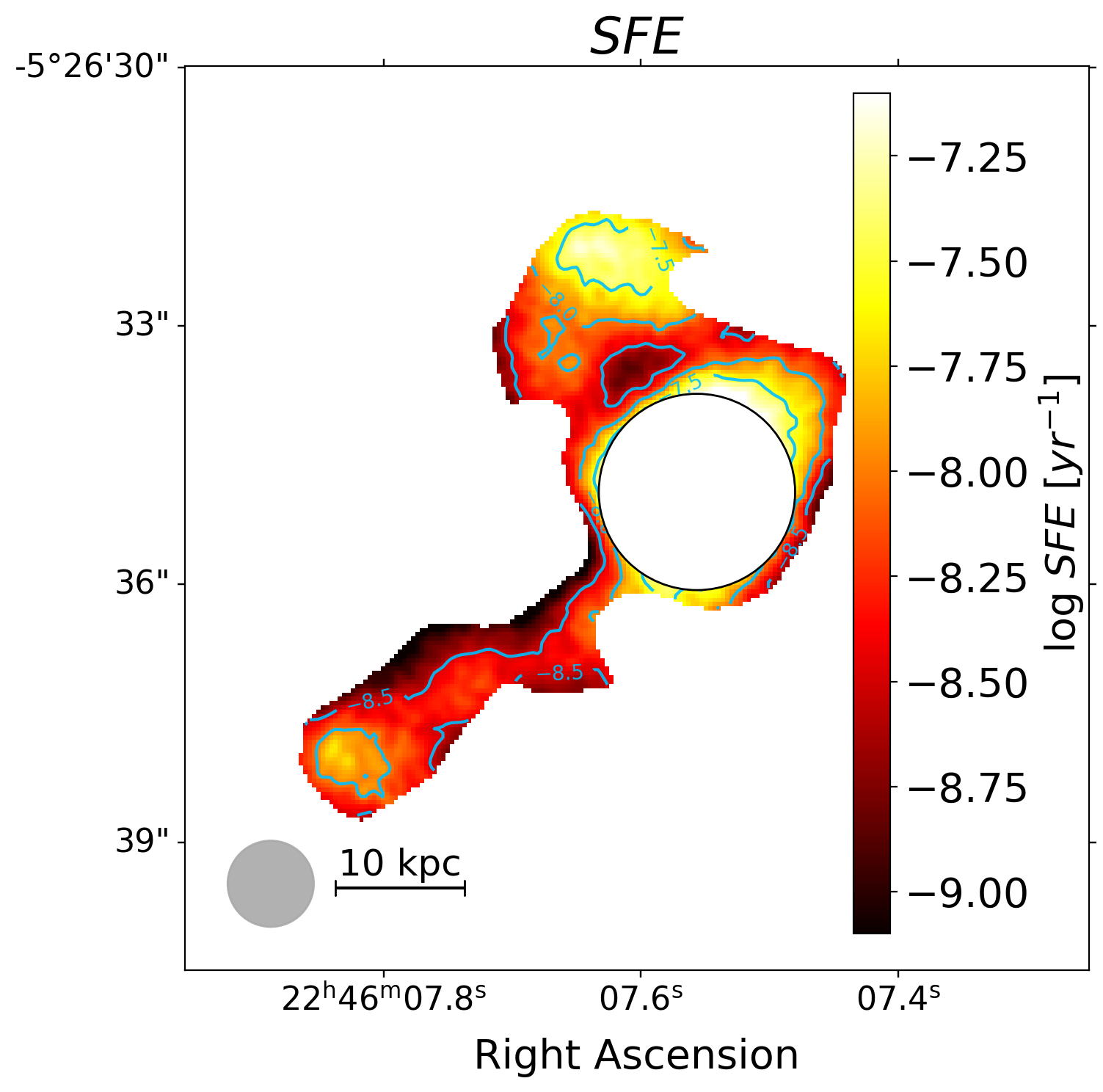}}
        \caption{Same as Fig.~\ref{fig:dustmaps} but for the derived dust properties (optical depth, infrared luminosity, star formation rate surface density ($\Sigma_{\mathrm{SFR}}$), and star formation efficiency (SFE)). The central $\sim$\,8~kpc radius is marked in the $\Sigma_{\mathrm{SFR}}$ map and masked in the SFE map, as it is likely affected mainly by quasar heating. Some areas in the system seem to have very high SFE, with corresponding depletion times of only a few tens of Myr.}
        \label{fig:dustmaps_derived}
\end{figure*}

Fig.~\ref{fig:dustmaps_derived} shows other derived dust properties. The optical thickness $\tau_{\nu}$ using Eq.~\ref{eq:3} is shown at rest-frame 158~${\upmu\mathrm{m}}$, and indicates that even at these wavelengths there are some areas in the system displaying significant absorption. We also present the total IR luminosity ($3-1100 \upmu\mathrm{m}$) map derived by integrating the fitted modified blackbody of each pixel. $L_{3-1100\upmu\mathrm{m}}$ peaks at $\sim6\times10^{13}\ L_{\odot}$~beam$^{-1}$ in the central region, and gradually descends to $\sim10^{11}-10^{12}\ L_{\odot}$. Additionally, we estimate the star formation rate (SFR) surface density ($\Sigma_{SFR}$) as $\Sigma_{SFR}=\mathrm{SFR}/A$. The SFR is estimated from $L_{IR}$ as in \cite{kennicutevans2012}, and $A$ is the intrinsic area derived from the modified blackbody fitting. Lastly, we show the star formation efficiency (SFE), calculated as $\mathrm{SFE=SFR}/M_{gas}$, with $M_{gas}=M_{d}\times\delta_{GDR}$. The SFE is enhanced in the north and the southeast of the system, reaching values as high as $10^{-7.25}$~yr$^{-1}$. 

\section{Discussion}
The temperature peak of $\sim$\,110~K measured in W2246--0526 is among the highest measured to date in galaxies at high-redshift, and is consistent with quasar heating. Integrating the modified blackbody fitted to the central beam results in a $L_{3-1100\upmu\mathrm{m}}=6.4\times10^{13}\ L_{\odot}$ (note that this does not include emission from hotter dust, which dominates the mid- and near-IR). To reproduce this luminosity we would require a $\Sigma_{SFR}\sim10^4\ M_{\odot}\ \mathrm{yr}^{-1}\ \mathrm{kpc}^{-2}$. This is well above the characteristic Eddington limit set by radiation pressure on dust grains in optically thick disks of $\Sigma_{SFR}\sim10^3\ M_{\odot}\ \mathrm{yr}^{-1}\ \mathrm{kpc}^{-2}$ \citep{thompson2005}, suggesting that an obscured AGN is a more likely source of heating than an obscured, compact, extreme starburst. This agrees with other studies stating that a starburst is unlikely to be the dominant source of energy in extremely luminous hot DOGs \citep{eisenhardt2012,tsai2015}. 

Fig.~\ref{fig:regions} shows that we resolve the dust emission over scales of 50~kpc. Theoretical models predict that sources with $L_{\mathrm{bol}}>10^{14}\ L_{\odot}$ can heat dust to temperatures of $30-40$~K at distances of tens of kpc if there are optically thin lines of sight \citep{scoville2013}. In such cases, a temperature gradient would be expected, with dust being colder farther from the central quasar. However, our results show a fairly constant dust temperature across the tidal streams and companion galaxies. While we cannot entirely rule out quasar heating, especially if optical depth variations are present along the quasar's line of sight, the uniformity of the temperatures suggests that the dust in these extended structures is predominantly heated by in-situ star formation. This indicates that stars are likely being formed within the dusty streams as they fly by or are accreted onto the central quasar and its host galaxy. We also estimate the SFR in the extended component to be $\sim800\ M_{\odot}\ \mathrm{yr}^{-1}$ by integrating the luminosity of the system outside the central 2~$\arcsec$ region displaying higher dust temperature. This is consistent with previous SFR estimations for W2246--0526 from total source integrated values \citep{sun2024}, and lower and upper limits for the companions \citep{diaz2018}. We speculate that the stellar mass being accreted into the central host during the merger process may include a contribution from star formation occurring in infalling, non-self-gravitating structures. In addition, a fraction of the stars formed in the dusty streams in the W2246--0526 system may end up as part of a future stellar halo. 

Fig.~\ref{fig:dustmaps} and Fig.~\ref{fig:dustmaps_derived} show central anisotropies in $A$, $\beta$, and $\tau_{\nu}$. The area parameter has significant uncertainties (i.e., $A=0.90_{-0.38}^{+0.60}$~kpc$^2$ for the central beam), which makes the observed asymmetry in $A$ tentative. The fit to the $\beta$ parameter is much more robust, however, and we also observe an anisotropy, with higher $\beta$ values toward the southwest. The optical depth map shows the same anisotropy, but it virtually disappears if we fix $\beta$ in the fit. Therefore, higher optical depths are mainly driven by higher values of $\beta$.

Lastly, the high SFE regions coincide with the position of two companion galaxies detected in \cite{diaz2018}, to the north and the southeast of the hot DOG (C2 and C3 in Fig.~\ref{fig:regions}). Their $\Sigma_{SFR}\sim10^2\ M_{\odot}\ \mathrm{yr}^{-1}\ \mathrm{kpc}^{-2}$ is consistent with star formation, while their SFE correspond to molecular gas depletion times of less than 100~Myr, a relatively low value compared with starburst galaxies at low and high redshifts \citep[e.g.,][]{huang2014,elbaz2018,tacconi2018}. Previous studies have suggested the SFE can be enhanced by ongoing mergers \citep[e.g.,][]{genzel2010,diazgarcia2020,sargent2024}. Our result agrees with this interpretation, since W2246--0526 is experiencing multiple mergers. Although we cannot rule out the presence of additional obscured AGN in these companions, their contribution to the dust heating is likely minor, as suggested by their uniform temperature compared to the streams.
\section{Summary}
\label{sec:summary}

We present spatially resolved ALMA multiband continuum observations of the W2246--0526 hot DOG merging system, at $z=4.6$. Fitting a modified blackbody to the data pixel-by-pixel, we derive for the first time the spatially-resolved dust properties over a scale of $\sim$\,50~kpc in a system only 1.3~Gyr after the Big Bang. The system contains an estimated total mass of dust of $5.1\times10^8M_{\odot}$. The dust temperature peaks at the position of the hot DOG at $\sim$\,110~K and decreases radially, suggesting dust is being heated by the central quasar up to a distance of $\sim$\,8~kpc. The dust in the tidal streams around the hot DOG and within the companion galaxies is at a relatively uniform temperature of 30--40~K with no evident gradients, suggesting the streams and companions are heated by obscured, recent star formation. These newly formed stars could contribute to the mass growth of the central galaxy and SMBH in W2246--0526 or eventually become part of a stellar halo, highlighting the potential role of extended structures in the mass assembly of high-redshift galaxies.

\begin{acknowledgements}
      This work is supported by the European Research Council under grant agreement No. 771282. MA acknowledges support from the ANID Basal Project FB210003 and ANID MILENIO NCN2024\_112. RJA was supported by FONDECYT grant number 1231718 and by the ANID BASAL project FB210003. C.-W. T. is Supported by NSFC No. 11988101 and the International Partnership Program of the Chinese Academy of Sciences, Program No. 114A11KYSB20210010. RD acknowledges support from the INAF GO 2022 grant “The birth of the giants: JWST sheds light on the build-up of quasars at cosmic dawn” and from PRIN MUR “2022935STW” RFF M4.C2.1.1, CUP J53D23001570006 and C53D23000950006. Portions of this research were carried out at the Jet Propulsion Laboratory, California Institute of Technology, under a contract with NASA. This paper makes use of the following ALMA data: ADS/JAO.ALMA\#2013.1.00576.S, ADS/JAO.ALMA\#2015.1.00883.S, ADS/JAO.ALMA\#2016.1.00668.S, ADS/JAO.ALMA\#2017.1.00899.S, ADS/JAO.ALMA\#2018.1.00119.S, 
      ADS/JAO.ALMA\#2018.1.00333.S,
      ADS/JAO.ALMA\#2021.1.00726.S. ALMA is a partnership of ESO (representing its member states), NSF (USA) and NINS (Japan), together with NRC (Canada), NSTC and ASIAA (Taiwan), and KASI (Republic of Korea), in cooperation with the Republic of Chile. The Joint ALMA Observatory is operated by ESO, AUI/NRAO and NAOJ.
\end{acknowledgements}

\bibliographystyle{aa}
\bibliography{refs} 

\begin{appendix}
\section{Intensity maps}\label{appendix}

\begin{figure*}
    \centering
        \subfloat{\includegraphics[width=85mm]{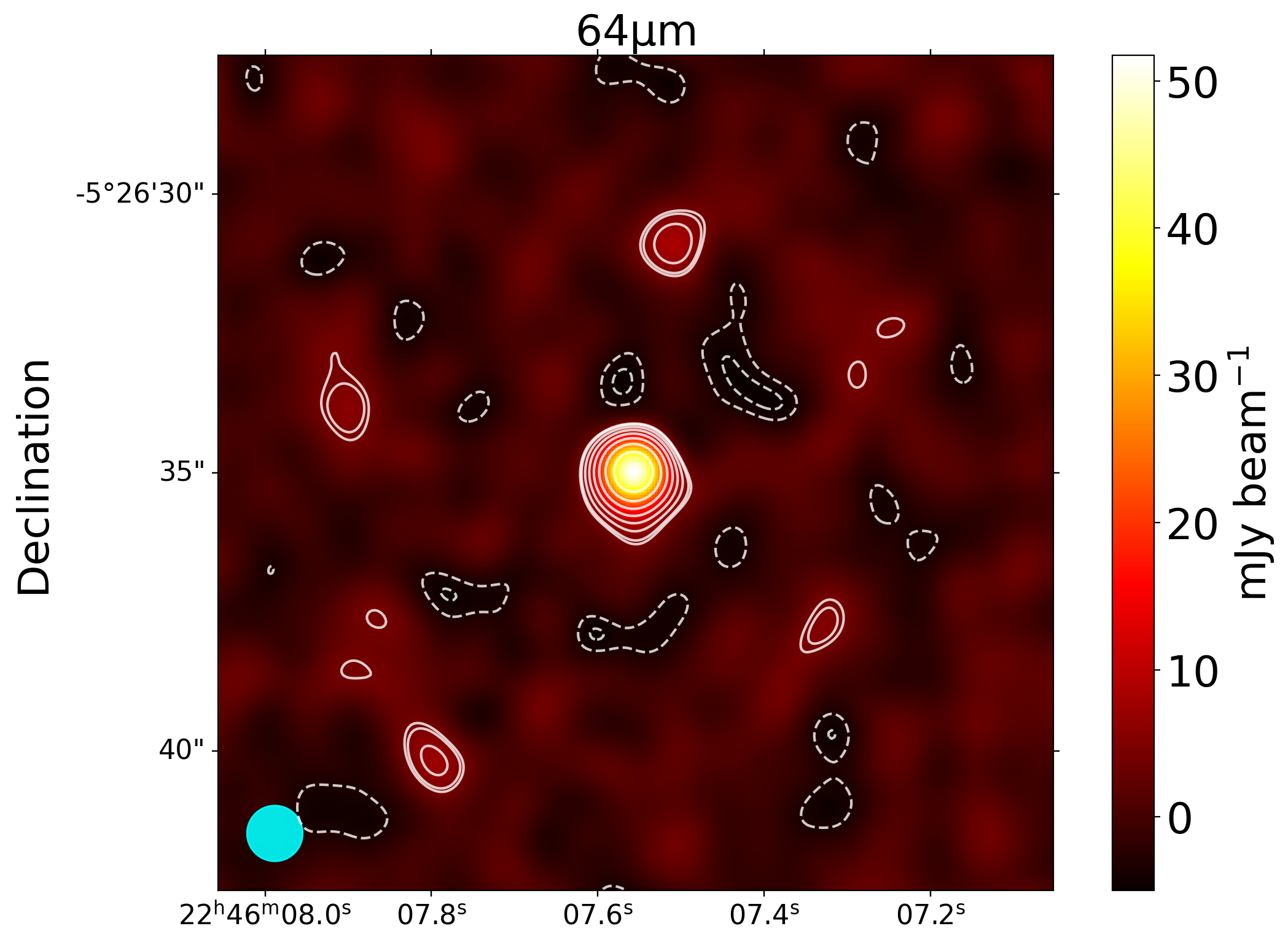}}
        \hspace{-2mm}
        \subfloat{\includegraphics[width=81mm]{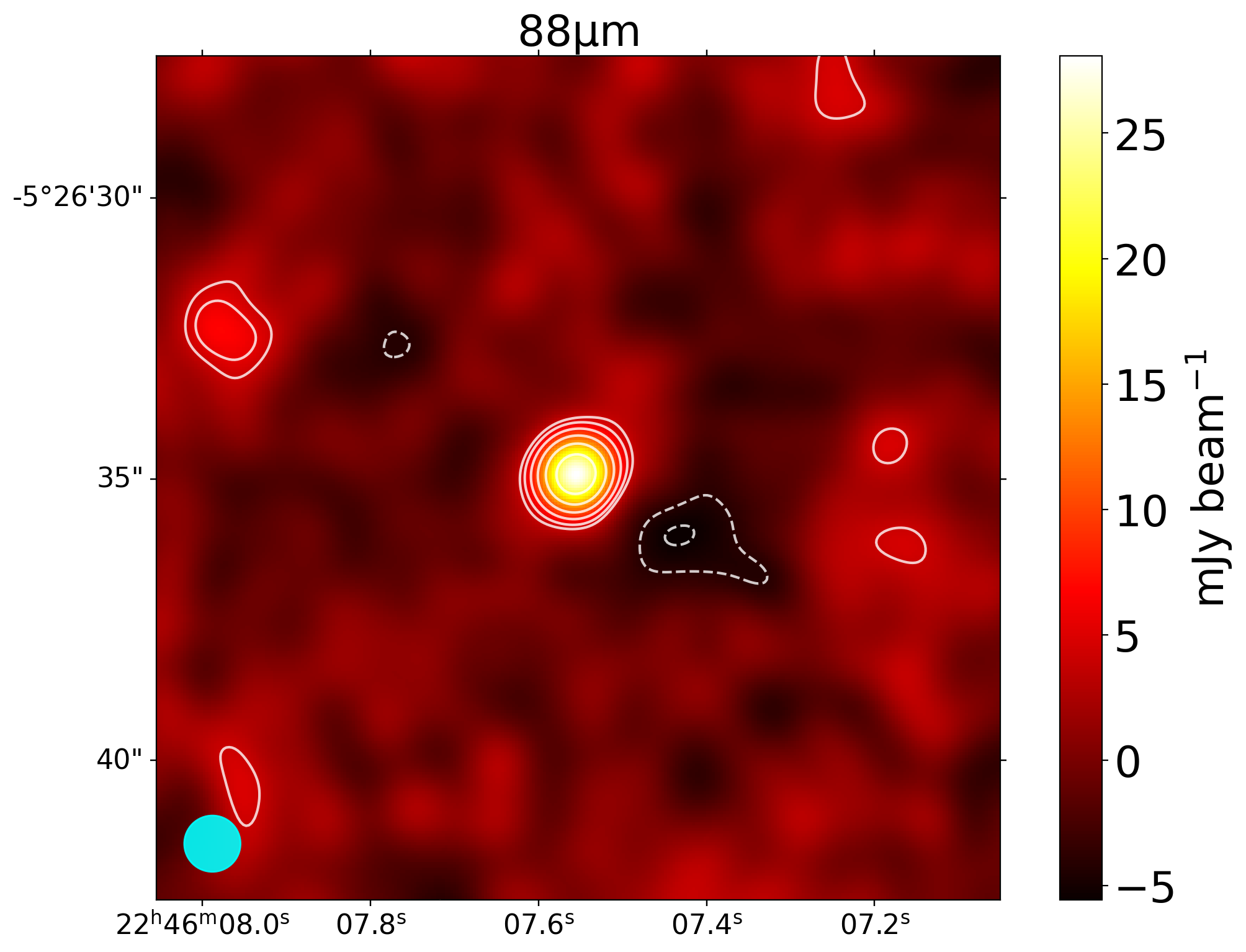}}
        \\
        \hspace{-3mm}
        \subfloat{\includegraphics[width=85mm]{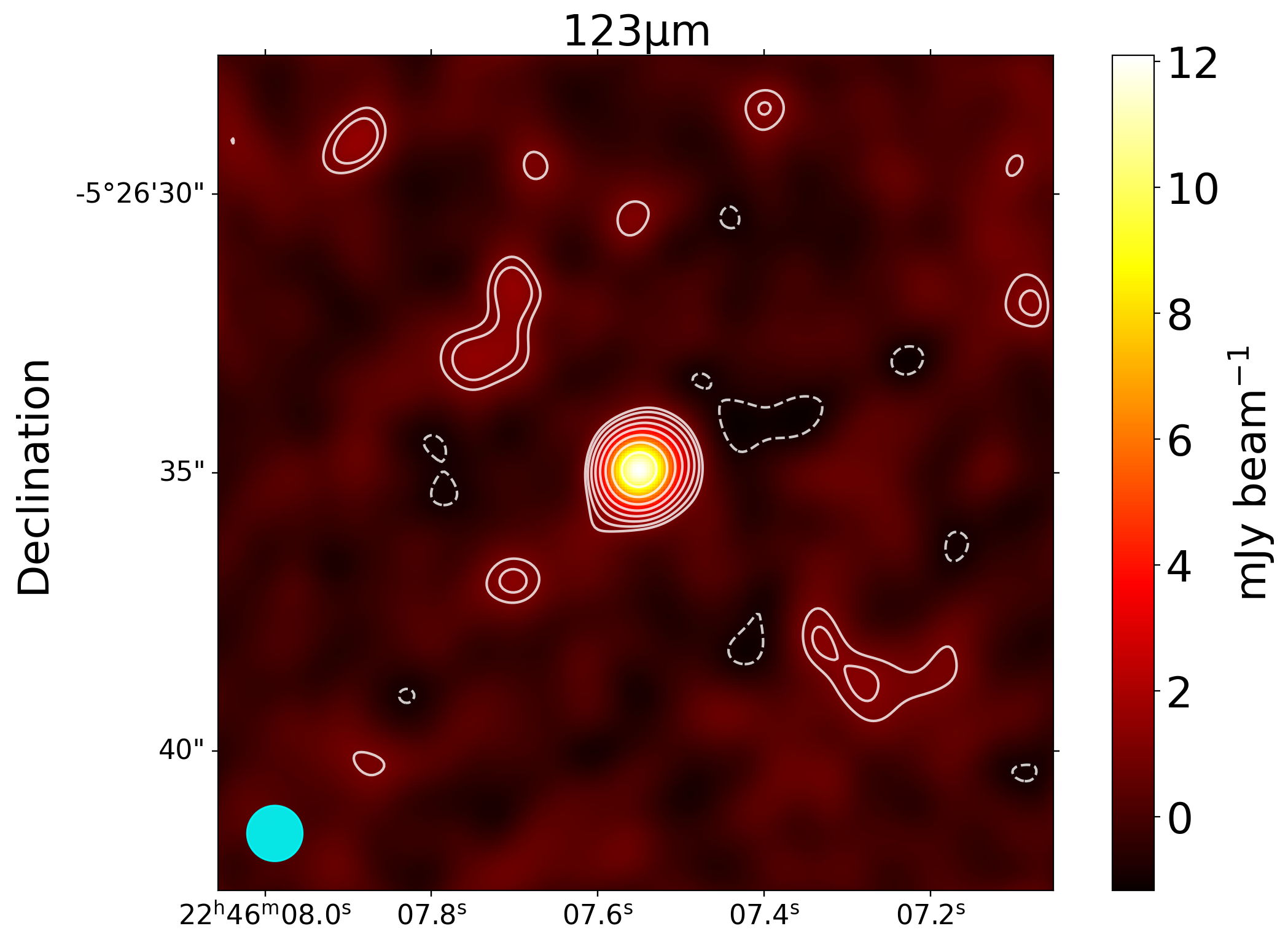}}
        \hspace{-2mm}
        \subfloat{\includegraphics[width=79mm]{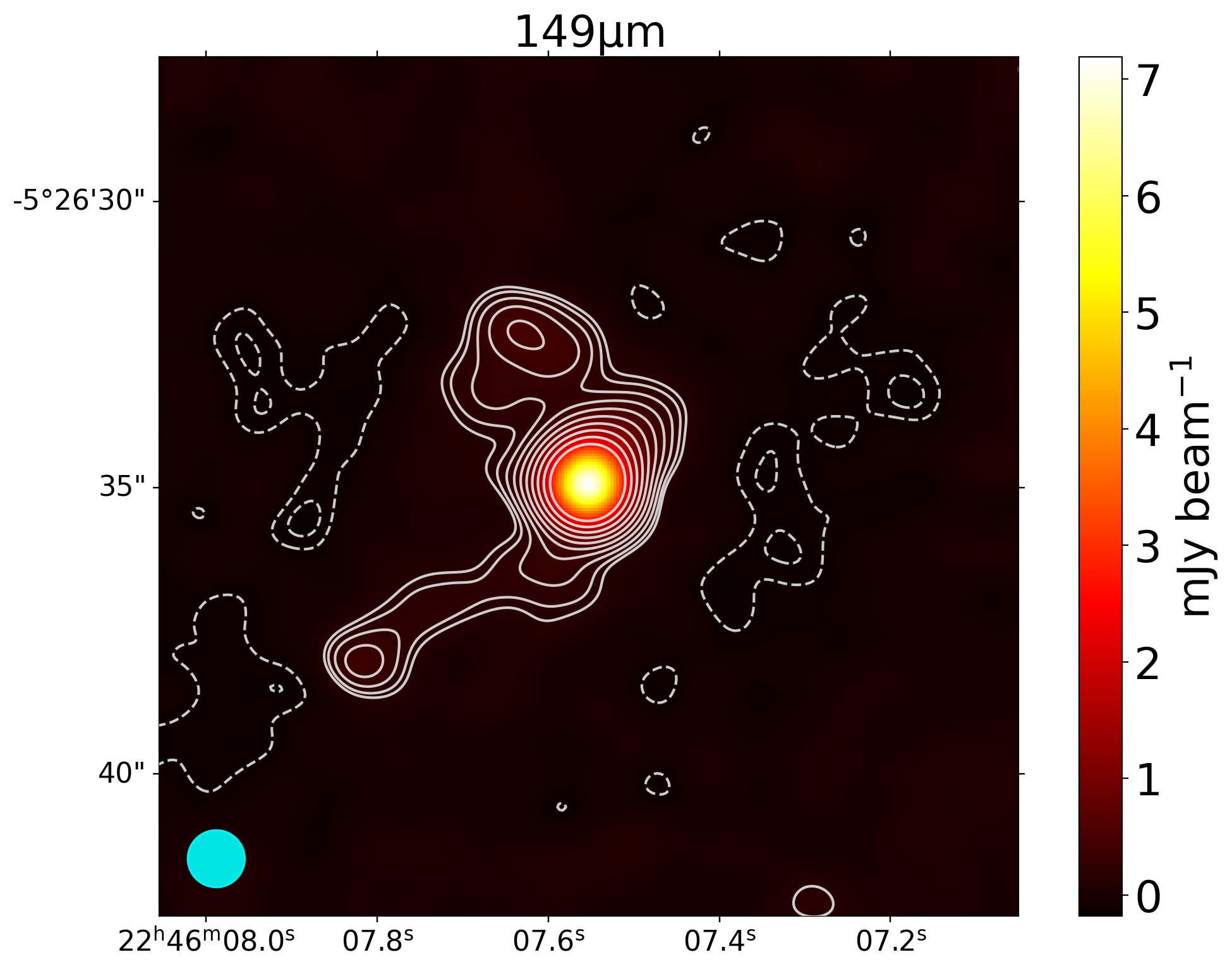}}
        \\
        \subfloat{\includegraphics[width=84mm]{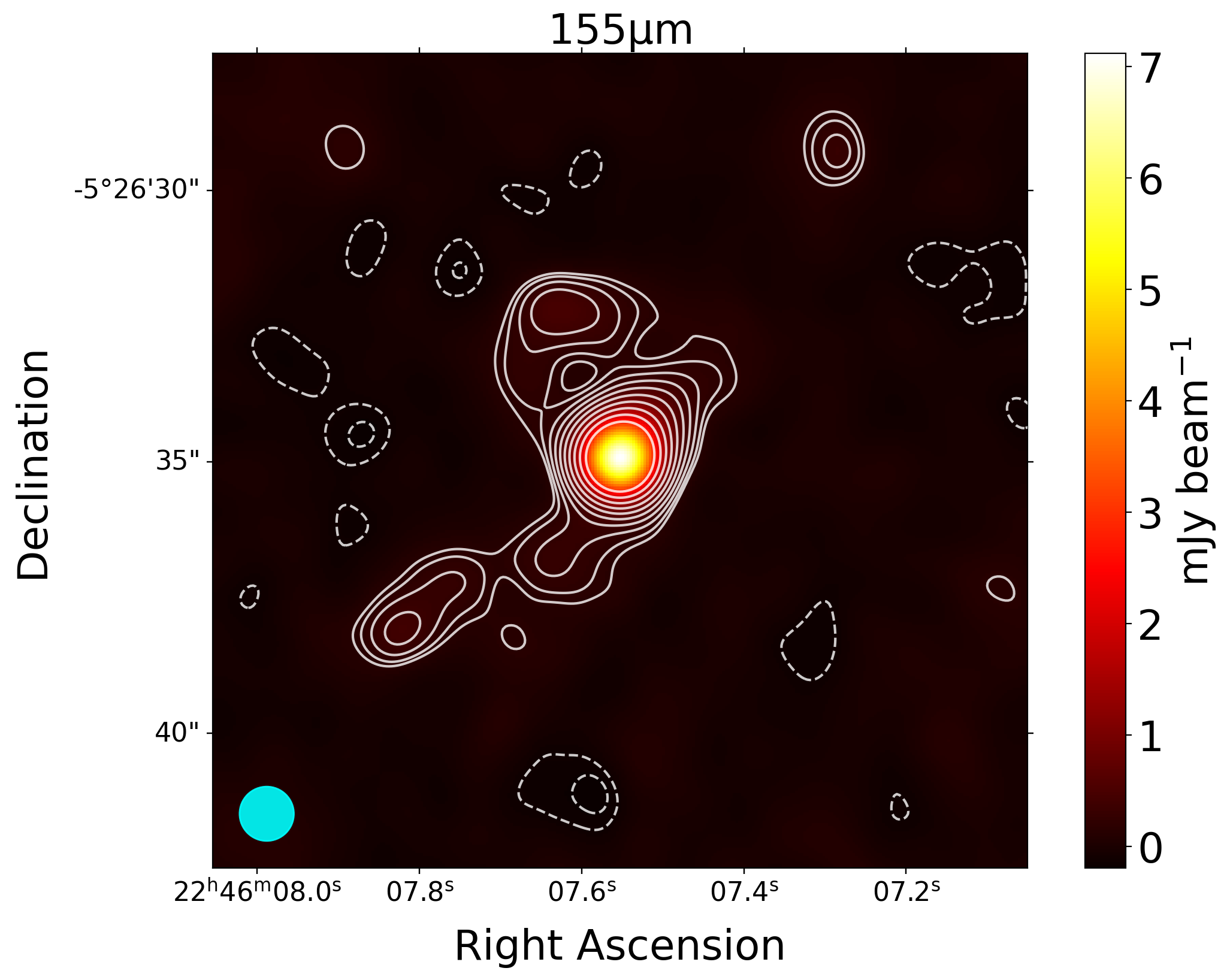}}
        \subfloat{\includegraphics[width=82mm]{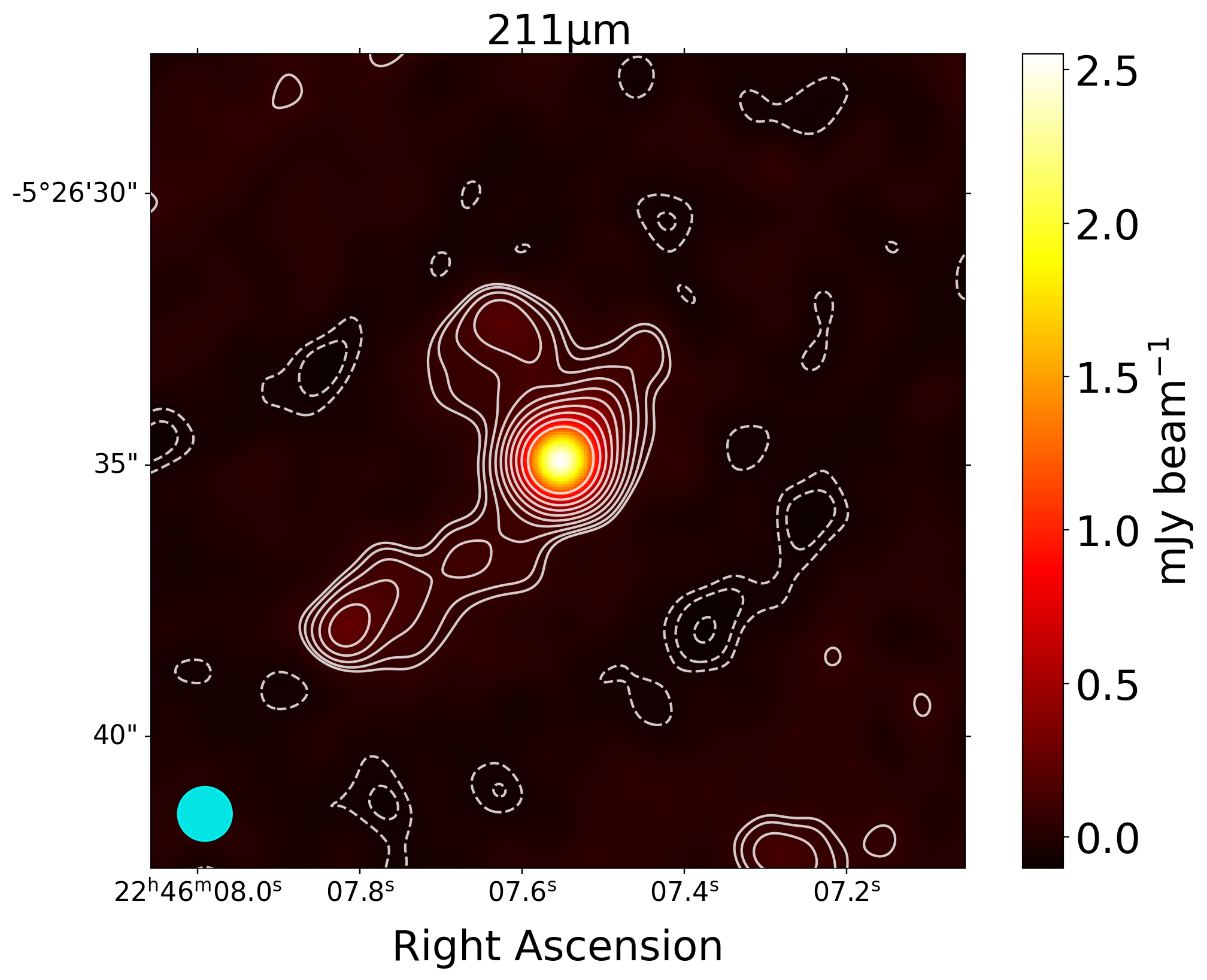}}
        \caption{Intensity maps for the datasets of the W2246--0526 merging system. The central rest-frame wavelength of each continuum dataset is shown at the top of each panel, and the cyan circle in the bottom-left of each panel represents the size of the ALMA beam (1$\arcsec$). White, solid contours indicate [3, 2$^{n/2}]\times\upsigma$ levels, with $\upsigma$ being the rms of the map, and $n$=[4,5,6,...]. Dashed contours show negative flux at the same absolute levels.}
        \label{fig:A_mom0}
\end{figure*}

\begin{figure*}[h!]
    \ContinuedFloat 
    \centering
    \subfloat{\includegraphics[width=87mm]{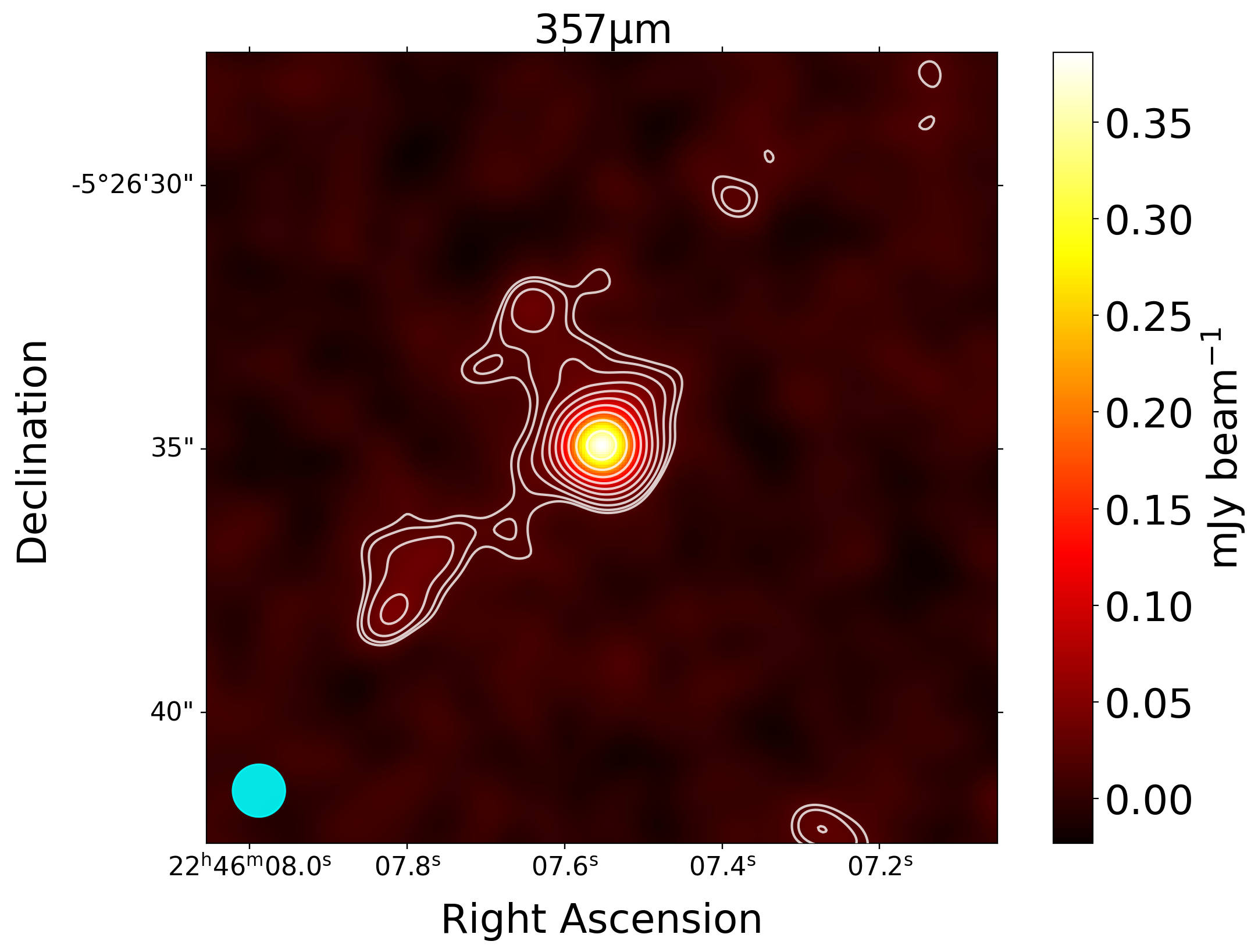}}
    \subfloat{\includegraphics[width=85mm]{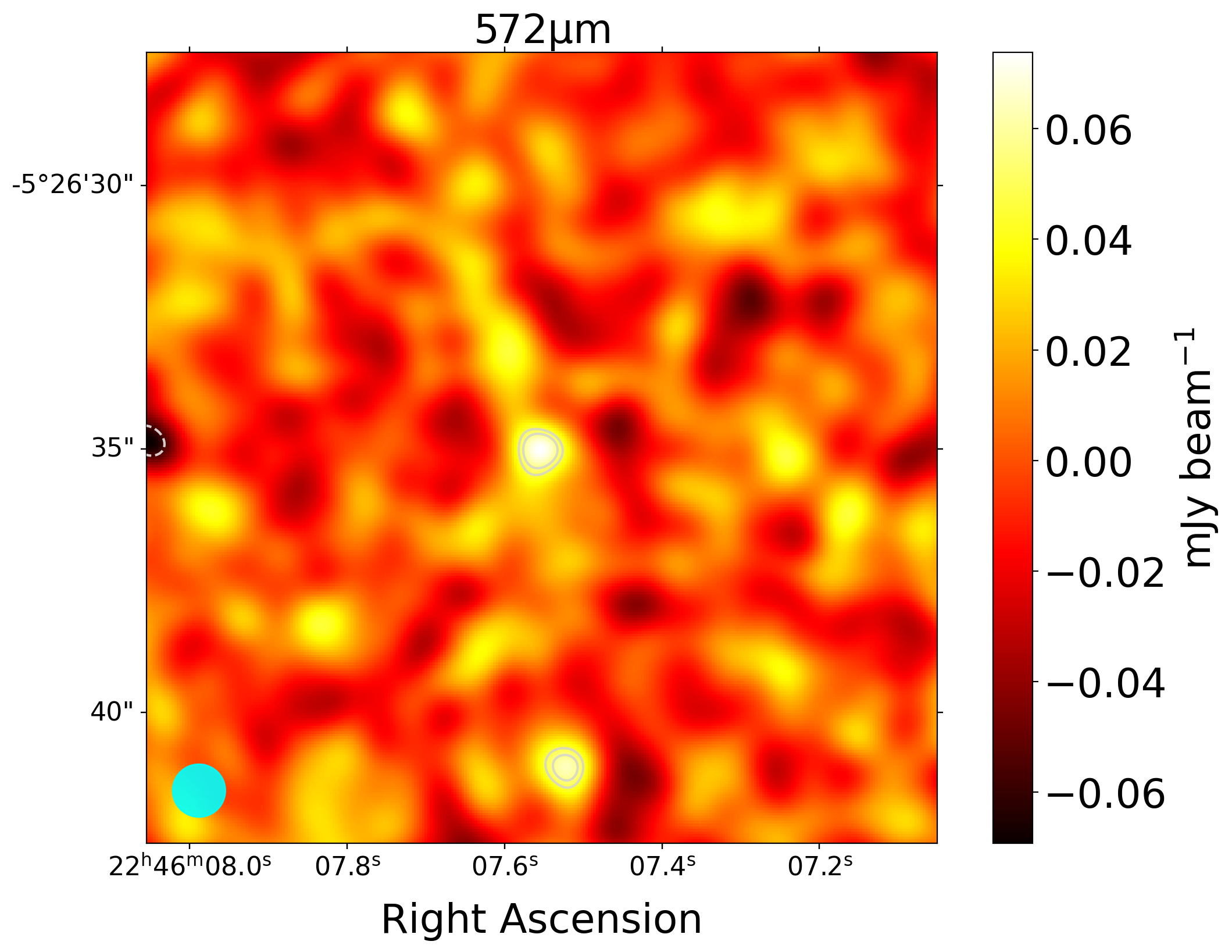}}
    \caption{Continuation.}
\end{figure*}

\end{appendix}

\end{document}